\documentclass[twocolumn,showpacs]{revtex4}

\usepackage{graphicx}
\usepackage{dcolumn}
\usepackage{amsmath}

\makeatletter
\def\btt#1{\texttt{\@backslashchar#1}}%
\DeclareRobustCommand\bblash{\btt{\@backslashchar}}%
\makeatother

\begin{document}
\pagestyle{empty}
\preprint{PrOsSb}

\title{Evidence for Unconventional Strong-coupling Superconductivity in PrOs$_4$Sb$_{12}$ : An Sb Nuclear Quadrupole Resonance (NQR) Study}
\author{H.~Kotegawa$^{1}$, M.~Yogi$^{1}$, Y.~Imamura$^{1}$, Y.~Kawasaki$^{1}$,  G.~-q.~Zheng$^{1}$, Y.~Kitaoka$^{1}$, S.~Ohsaki$^{2}$, H.~Sugawara$^{2}$, Y.~Aoki$^{2}$, and H.~Sato$^{2}$}

\address{$^1$Department of Physical Science, Graduate School of Engineering Science, Osaka University, Toyonaka, Osaka 560-8531, Japan\\
$^{2}$ Graduate School of Science, Tokyo Metropolitan University, Minami-Ohsawa 1-1, Hachioji, Tokyo 192-0397, Japan}

\date{\today}

\begin{abstract}

We report Sb-NQR results which evidence a heavy-fermion (HF) behavior and an unconventional superconducting (SC) property in the filled-skutterudite compound PrOs$_4$Sb$_{12}$ revealing a SC transition temperature $T_c=1.85$ K. The temperature ($T$) dependence of nuclear-spin-lattice-relaxation rate $1/T_1$ and NQR frequency unravel a low-lying crystal-electric-field splitting below $T_0\sim 10$ K, associated with Pr$^{3+}$($4f^2$)-derived ground state. The emergence of $T_1T$=const. behavior below $T_F\sim 4$ K points to the formation of heavy-quasiparticle state. In the SC state,  $1/T_1$ shows neither a coherence peak nor a  $T^3$like power-law behavior observed for HF superconductors to date. The isotropic energy-gap with a size of gap $\Delta/k_B=4.8$ K begins to already open up at $T^*\sim$ 2.3 K without any coherence effect just below $T_c=1.85$ K. We highlight that the superconductivity in PrOs$_4$Sb$_{12}$, which is in an unconventional strong-coupling regime, differs from a conventional s-wave type and any unconventional ones with the line-node gap.
\end{abstract}

\vspace*{5mm}

\maketitle

A class of Ce- and U-based intermetallic compounds reveal a crossover from high-temperature localized to low-temperature heavy-fermion (HF) state in which $f$ electrons are delocalized with enormous effective mass, and are remarkable to undergo an unconventional superconducting (SC) transition with line-node gap, indicative of the Cooper pairing of heavy fermions with an angular momentum greater than zero. It is widely believed that the superconductivity in these materials is mediated by magnetic fluctuations. 

Recently, the HF-like behavior or a quadrupolar ordering has been reported in PrInAg$_2$ \cite{Yatskar} and PrFe$_4$P$_{12}$ \cite{Sugawara,Matsuda,Aoki,Nakanishi} and PrPb$_3$ \cite{Bucher,Morin,DAoki,Tayama}. In these compounds, the ground state of a Pr$^{3+}$ with $4f^2$ in a crystal electric field (CEF) scheme is believed to be the $\Gamma_3$ nonmagnetic doublet with the electric quadrupolar moments for a total angular momentum $J=4$ state. The quadrupolar moments interact with the charges of conduction electrons, leading to the HF-like behavior or the quadrupolar ordering in these Pr-based compounds. In fact, PrPb$_3$ shows an antiferro-quadrupolar ordering in the $\Gamma_3$ ground state at 0.4 K \cite{Bucher,Morin,DAoki,Tayama}. In PrInAg$_2$, a broad peak in specific heat is identified as a Kondo anomaly, having an enhanced electronic specific heat coefficient $\gamma\sim 6.5~{\rm J/mol K^2}$ \cite{Yatskar}. Likewise, the filled-skutterudite compound PrFe$_4$P$_{12}$ shows the HF-like behavior with a large mass of $m^*\sim 70~m_e$ \cite{Sugawara} under a magnetic field and undergoes an anomalous transition at 6.4 K at zero field, indicative of a quadrupolar ordering \cite{Matsuda,Aoki,Nakanishi}. Note that the large values in $C/T$ for these compounds  are not always ascribed to such low-energy degrees of freedom as either magnetic or quadrupolar fluctuations, but also to the Schottky anomaly originating from some low lying CEF splitting.

Meanwhile, Bauer {\it et al.} reported the observation of HF behavior and superconductivity at $T_c=1.85$ K in the filled-skutterudite compound PrOs$_4$Sb$_{12}$  that is the first case of Pr-based HF superconductor \cite{Bauer}. 
Its HF state  was inferred from the jump in the specific heat at $T_c$, the slope of the upper critical field near $T_c$, and the electronic specific heat coefficient $\gamma\sim 350-500$ mJ/mole K$^2$.  The magnetic susceptibility, thermodynamic measurements, and recent inelastic neutron scattering experiments revealed the ground state of the Pr$^{3+}$ ions in the cubic CEF to be the $\Gamma_3$ nonmagnetic doublet \cite{Bauer,Woodward}.
In the Pr-based compounds with the $\Gamma_3$ ground state, the quadrupolar interactions play important role. In analogy with the quadrupolar Kondo model \cite{Cox}, it was suggested that the HF-like behavior exhibited by PrOs$_4$Sb$_{12}$ may be relevant to a quadrupolar Kondo lattice. An interesting issue to be addressed is what role of Pr$^{3+}$-derived  quadrupolar fluctuations are relevant with the onset for the superconductivity in this compound.

In this letter, we report the observation of unconventional SC property probed by the measurement of nuclear-spin-lattice-relaxation time $T_1$ in PrOs$_4$Sb$_{12}$ through  $^{121,123}$Sb nuclear quadrupolar resonance (NQR) experiments at zero field. 
Single crystals of PrOs$_4$Sb$_{12}$ were grown by the Sb-flux method as described elsewhere \cite{Sugawara2}.
Measurements of electrical resistivity and ac-susceptibility confirmed a superconducting (SC) transition at $T_c=1.85$ K. 
The observation of the de Haas-van Alphen (dHvA) oscillations ensures high-quality of the samples \cite{Sugawara2}.
For the $^{121,123}$Sb NQR measurements, the single crystals were crushed into powder.

Fig.~1(a) displays $^{121,123}$Sb-NQR spectra for two Sb isotopes at 4.2 K. Note that $^{121}$Sb($^{123}$Sb) with natural abundance 57.3 (42.7) \% has the nuclear spin $I=5/2$ (7/2) and the nuclear gyromagnetic ratio $\gamma_N = 10.189~(5.5175)$ [MHz/T], and exhibits two (three) NQR transitions. The ratio of nuclear quadrupolar moment, $^{123}Q/^{121}Q$ was reported as $\sim$1.275 from the NQR measurement of the pure Sb metal \cite{Hewitt}. From these spectra, the values of nuclear quadrupole frequency, $^{121}\nu_Q\sim$ 44.2 MHz and $^{123}\nu_Q\sim$ 26.8 MHz, are deduced for PrOs$_4$Sb$_{12}$ along with an asymmetric parameter $\eta\sim 0.46$. Note that the nuclear electric quadrupole interaction is represented as
\begin{eqnarray}
{\cal H}_Q=\frac{e^2qQ}{4I(2I-1)}\{[3I_z^2-I(I+1)]+\eta(I_x^2-I_y^2)\}, 
\end{eqnarray}
where $eq$ gives the component along the principle $z$-axis of electric field gradient (EFG), which is determined by the charge distribution of conduction electrons around the Sb nuclei.

Fig.~1(b) indicates the $T$ dependence of the $^{123}$Sb-NQR spectrum arising from the 2$\nu_Q$ transition where $\nu_Q$ is defined as $\frac{e^2qQ}{4I(2I-1)}$. A peak in the spectrum shifts to a high frequency below $T_0\sim$ 10 K that is shown later to be a characteristic temperature in $1/T_1$. 
Since the transitions of $2\nu_Q$ and $3\nu_Q$ are not sensitive to the change 
in $\eta$, almost the same relative variation of EFG, $[q-q_0]/q_0$ against the value $q_0$ at 10 K is obtained from the $2\nu_Q$ for $^{123}$Sb and the $2\nu_Q$ and $3\nu_Q$ for $^{121}$Sb, using a relation of $[\nu_Q(T)-\nu_0]/\nu_0$ as shown in Fig.~1(c). Here $\nu_0$ is the value of $\nu_Q$ at 10 K.
The rapid increase in EFG at the Sb site is evident below $T_0\sim10$ K.
Such the distinct shift is never seen in the isostructual LaOs$_4$Sb$_{12}$.
It is natural to ascribe this increase in EFG  to the Pr$^{3+}$-derived change. This is because the electronic contribution in specific heat divided by $T$, $\Delta C/T$ revealed a similar $T$ variation to the EFG. It was reported that the rapid increase in $\Delta C/T$ is consistent with the energy scheme of the $4f^{2}(J=4)$ state of Pr$^{3+}$ in the CEF, that is, the $\Gamma_3$ nonmagnetic doublet is a ground state and the $\Gamma_5$ magnetic triplet is first excited state. The magnetic susceptibility, inelastic neutron scattering, and specific heat revealed  the CEF energy splitting of $\Delta_{CEF}=7 \sim 11$ K between these two levels \cite{Bauer,Woodward,Aoki2,Vollmer}. We remark, therefore, that the electric quadrupole moments of this $\Gamma_3$ ground state interact with the charges of the conduction electrons in the $T$ range below $T_0\sim$ 10 K because of $T<\Delta_{CEF}$. Thus the significant increase below $T_0$ in both $\Delta C/T$ and the EFG at the Sb site is indicative of a low-lying CEF splitting below $T_0\sim 10$ K, associated with Pr$^{3+}$($4f^2$)-derived ground state.

Fig.~2 indicates the $T$ dependence of $1/T_1$ measured at $2\nu_Q\sim 48.9$ MHz for $^{123}$Sb along with the result in LaOs$_4$Sb$_{12}$ ($T_c=0.75$ K). In the normal state, a relation of $T_1T =$ const. is valid in LaOs$_4$Sb$_{12}$, characteristic for conventional metallic materials. By contrast, the $1/T_1$ at the normal state for PrOs$_4$Sb$_{12}$ is strongly enhanced than for LaOs$_4$Sb$_{12}$, showing a relaxation behavior similar to Ce-based HF systems reported thus far. Since $1/T_1$ stays a constant in $T=$10-20 K, the $4f$-electron-derived moments behave as if localized.  
With decreasing $T$ below $T_0\sim 10$ K, $1/T_1$ seems to be decreased because of $T<\Delta_{CEF}$. In the localized regime at high $T$, note $1/T_1T\propto \chi(T)$ where $\chi(T)$ is a Curie-Weiss-like magnetic susceptibility in the normal state. As a matter of fact, the $T$ dependence of $1/T_1T$, which is shown in Fig.~3, resembles the measured susceptibility, which suggests that the $\Gamma_3$ is the ground state and the $\Gamma_5$ is the first excited state \cite{Bauer}.
The relaxation behavior is thus consistent with the other experiments suggesting $\Delta_{CEF}=7 \sim 11$ K \cite{Bauer,Woodward,Aoki2,Vollmer}.
In $T$ lower than $\sim$ 4 K, however, this CEF model is not valid. 

The inset presents the $T$ dependencies of $(1/T_1)/\gamma_N^2$ at the $2\nu_Q\sim 48.9$ MHz and the $3\nu_Q\sim 78.6$ MHz for $^{123}$Sb, and the $2\nu_Q\sim 85.0$ MHz for $^{121}$Sb. All the data are consistently on a curve, demonstrating that the relaxation process is magnetic in origin throughout the measured $T$ range. This result indicates that the ground state is not always in a nonmagnetic regime where the quadrupolar degree of freedom of nonmagnetic doublet $\Gamma_3$ becomes dominant, but it might be hybridized with conduction electrons, making magnetic relaxation channel open even for $T<<\Delta_{CEF}$.  In such the case, 
 the relaxation process at low $T$ may be described in terms of the CEF channel and the quasiparticle's one as follows,
\begin{eqnarray}
1/T_1 = A \times \exp(-\Delta_{CEF}/k_BT)+B \times 0.7T
\end{eqnarray}
where $0.7$ is the value of $1/T_1T \sim$ const. for LaOs$_4$Sb$_{12}$.
The first term is the CEF contribution arising from the first excited $\Gamma_5$ triplet state and the second one the quasiparticle's one due to the hybridization between the $\Gamma_3$ state and conduction electrons.
The dotted line in the inset of Fig.~3 is a best fit obtained by assuming $\Delta_{CEF}/k_B=8$ K \cite{Aoki2} and $B=26.7$ below $T_F\sim$ 4 K. The latter value allows us to estimate that the effective density of state (DOS) for PrOs$_4$Sb$_{12}$ is $\sim 5.2$ times larger than that for LaOs$_4$Sb$_{12}$, because $1/T_1T$ is proportional to the square of the DOS.
This value is quite consistent with the result of $\gamma=313-350$ mJ/mol K$^2$ for PrOs$_4$Sb$_{12}$ \cite{Bauer,Vollmer} being five times larger than $\gamma=56$ mJ/mol K$^2$ for LaOs$_4$Sb$_{12}$ \cite{Aoki3}. It is expected that
the heavy-quasiparticle state is realized through mixing between the nonmagnetic $\Gamma_3$ doublet state and conduction electrons.

Next we deal with the SC property. Fig.~4(a) shows the $T$ dependence of the quasiparticle part $(1/T_1T)_{qp}$ which is the second term in Eq.~(2). Here the data are normalized by the value of $(1/T_1T)_{qp}$=const. in $T=2.3 - 4.2$ K.
$(1/T_1T)_{qp}$ does not show any coherence peak just below $T_c=1.85$ K, and yet starts to be already decreased below $T^*$=2.3 K far above 
$T_c$. Note that this $T^*$ is close to the temperature at which $C/T$ has a peak \cite{Vollmer}. These results point to the unconventional nature of the superconductivity in PrOs$_4$Sb$_{12}$.  Most Ce- or U-based HF superconductors show the power-law behavior of $1/T_1\sim T^3$ at low temperatures, consistent with the line-node SC gap \cite{Kitaoka,Mac}.
However, the $T$ dependence of $(1/T_1T)_{qp}$ in PrOs$_4$Sb$_{12}$ does not show a $T^3$ dependence ($T^2$ dependence in $1/T_1T$) as indicated by the solid line in Fig.~4(a). Instead, as presented in Fig.~4(b), it follows an exponential decrease with $\Delta/k_B=4.8$ K over {\it three orders of magnitude} in  $T^*\sim 2.3$ K$-0.3T_c$ across $T_c=1.85$ K. Recent $\mu$SR experiment also suggests an isotropic energy gap \cite{MacLaughlin}. It is unconventional that the isotropic SC gap for PrOs$_4$Sb$_{12}$ begins to already open up below $T^*\sim 2.3$ K far above $T_c=1.85$ K. Unconventional strong-coupling effect seems to give rise to preformed pairs around $T^*$ before a bulk SC transition takes place. This also seems to be relevant with the observation of the large jump $\Delta C/\gamma T_c\sim 3$ in the specific heat at $T_c$ \cite{Vollmer}. Apparently, the anomalous relaxation behavior in  PrOs$_4$Sb$_{12}$ contrasts with a conventional one for an s-wave case that is actually seen in the $T$ dependence of $1/T_1$ for LaOs$_4$Sb$_{12}$ with $T_c=0.75$ K in Fig.~2. Remarkably, in this SC state, $1/T_1$ shows 
the large coherence peak just below $T_c$, followed by the exponential dependence with the gap size of $2\Delta/k_BT_c\sim 3.2$ at low $T$. This 
clearly evidences that LaOs$_4$Sb$_{12}$ is the conventional weak-coupling BCS $s$-wave superconductor.

Although, in a strong-coupling regime, a significant suppression 
in the coherence peak was reported \cite{Ohsugi}, it is unexpected that the gap emerges below $T^*\sim 2.3$ K that is far above $T_c=1.85$ K in PrOs$_4$Sb$_{12}$. Although the recent thermal-conductivity experiment suggests  point-nodes 
gap \cite{Izawa}, evidence for the existence of any node in the SC gap is absent down to 0.3$T_c$. Unexpectedly, $1/T_1$ shows a saturation below 0.3$T_c$, remaining issue to conclude any precise argument on some gap structure. It is, however, evident as unraveled in this work that the novel superconductivity takes place under such the situation that the quadrupolar degree of freedom plays vital role for the formation of quasiparticle at temperatures lower than $\Delta_{CEF}=7\sim 11$ K. We also remark that $T_c=1.85$ K for PrOs$_4$Sb$_{12}$ is more enhanced than $T_c$=0.75 K for LaOs$_4$Sb$_{12}$ and furthermore, the Pr-based superconductor PrRu$_4$Sb$_{12}$ ($T_c$=1.3 K), which is characterized by a singlet CEF ground state, is the weak coupling s-wave superconductor with $2\Delta/k_BT_c\sim 3.1$ \cite{Yogi}. Therefore, it raises a question what type of pairing interaction is possible in mediating the Cooper pair to cause the unconventional strong-coupling superconductivity in PrOs$_4$Sb$_{12}$.

In summary, the temperature dependence of $1/T_1T$ in the new HF superconductor PrOs$_4$Sb$_{12}$, which is well scaled to the magnetic susceptibility, have revealed that the $4f$-derived moments behave as if localized at the higher $T$ than $T_0\sim$ 10 K. At temperatures below $T_0$,  the marked increase in NQR frequency at the Sb site and the decrease in $1/T_1$ unraveled a low-lying crystal-electric-field splitting, associated with Pr$^{3+}$($4f^2$)-derived ground state. These results are consistent with other experiments that have led to an estimation of  $\Delta_{CEF}=7\sim 11$ between the first excited $\Gamma_5$ magnetic triplet and the $\Gamma_3$ nonmagnetic doublet ground state.
In the lower $T$  than $T_F\sim$ 4 K, the relaxation process is well accounted for by incorporating both the CEF contribution arising  from the first excited $\Gamma_5$ triplet state and the $T_1T$=const. contribution from the heavy quasiparticle state. The latter is due to the hybridization of the nonmagnetic CEF doublet state with conduction electrons, that is, responsible for the onset of the superconductivity.

In the SC state,  $1/T_1$ shows neither the coherence peak nor the $T^3$ like power-law behavior observed for the HF superconductors to date. Rather, the isotropic energy-gap model with $\Delta/k_B=4.8$ K is consistent with the data below $T^*\sim$ 2.3 K across $T_c=1.85$ K. Most remarkably, the novel superconductivity in PrOs$_4$Sb$_{12}$ differs from the conventional s-wave type and any unconventional ones with the line-node gap. The opening of the large isotropic SC gap 2$\Delta/k_BT_c\sim 5.2$ suggests the unconventional strong-coupling effect due to the pairing formation along with the lack for any anomaly across $T_c$. Importantly, the present work provides  clue to gain insight into the $4f^2$-based superconductivity in PrOs$_4$Sb$_{12}$.  We believe that the HF superconductor PrOs$_4$Sb$_{12}$ is in a new type of unconventional strong-coupling regime, closely relevant to the quadrupolar degree of freedom. The present work hence opens a new research field of superconductivity in strongly correlated electron systems.

We are grateful for fruitful discussions with Kazumasa Miyake and Hisatomo Harima on $f^2$-based HF systems. One of authors (H.K.) thanks Y.~Tokunaga,@and  K.~Ishida for valuable discussions and comments. This work was supported by the COE Research (10CE2004) in a Grant-in-Aid for Scientific Research from the Ministry of Education, Sports, Science and Culture of Japan. 
One of authors (H.K.) has been supported by {\it JSPS Research Fellowships for Young Scientists}.

\clearpage
\begin{figure}[htbp]
\begin{center}
\includegraphics[width=.9\linewidth]{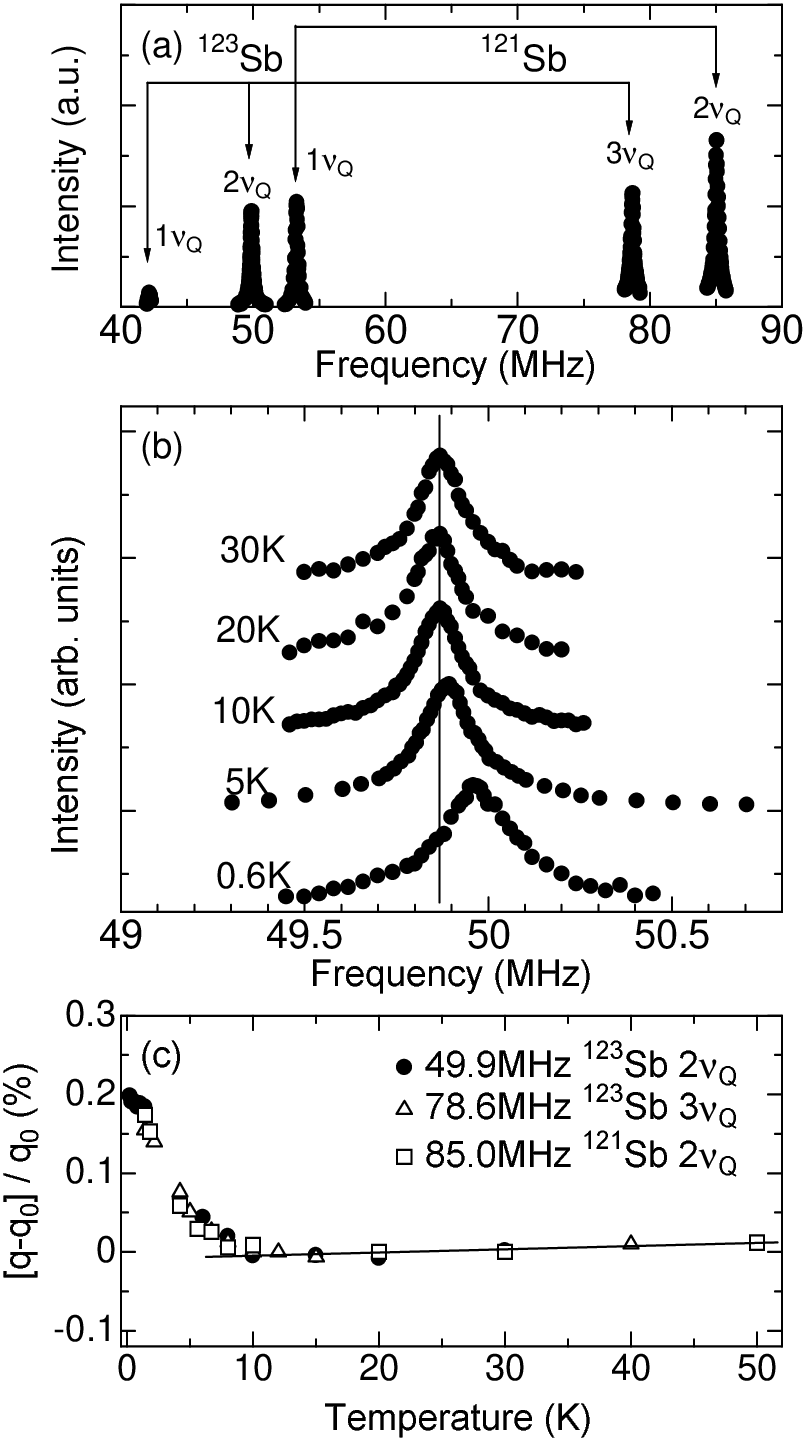}
\caption[]{\protect (a) $^{121}$Sb and $^{123}$Sb-NQR spectra in PrOs$_4$Sb$_{12}$. The respective electric quadrupole frequencies are estimated as $^{121}\nu_Q\sim$ 44.2 MHz and $^{123}\nu_Q\sim$ 26.8 MHz, and an asymmetric parameter as $\eta\sim 0.46$. (b) $T$ dependence of the NQR spectrum for the $2\nu_Q$ of $^{123}$Sb. The peak in the spectrum shifts significantly to the high 
frequency upon cooling below $\sim 10$ K. (c) $T$ dependence of the relative change in the electric field gradient (EFG), $[q(T)-q_0]/q_0$ at the Sb nuclei. Here $[q(T)-q_0]/q_0=[\nu_Q(T)-\nu_0]/\nu_0$ with the $\nu_0$ at 10 K.
}
\end{center}
\end{figure}

\clearpage
\begin{figure}[htbp]
\begin{center}
\includegraphics[width=.8\linewidth]{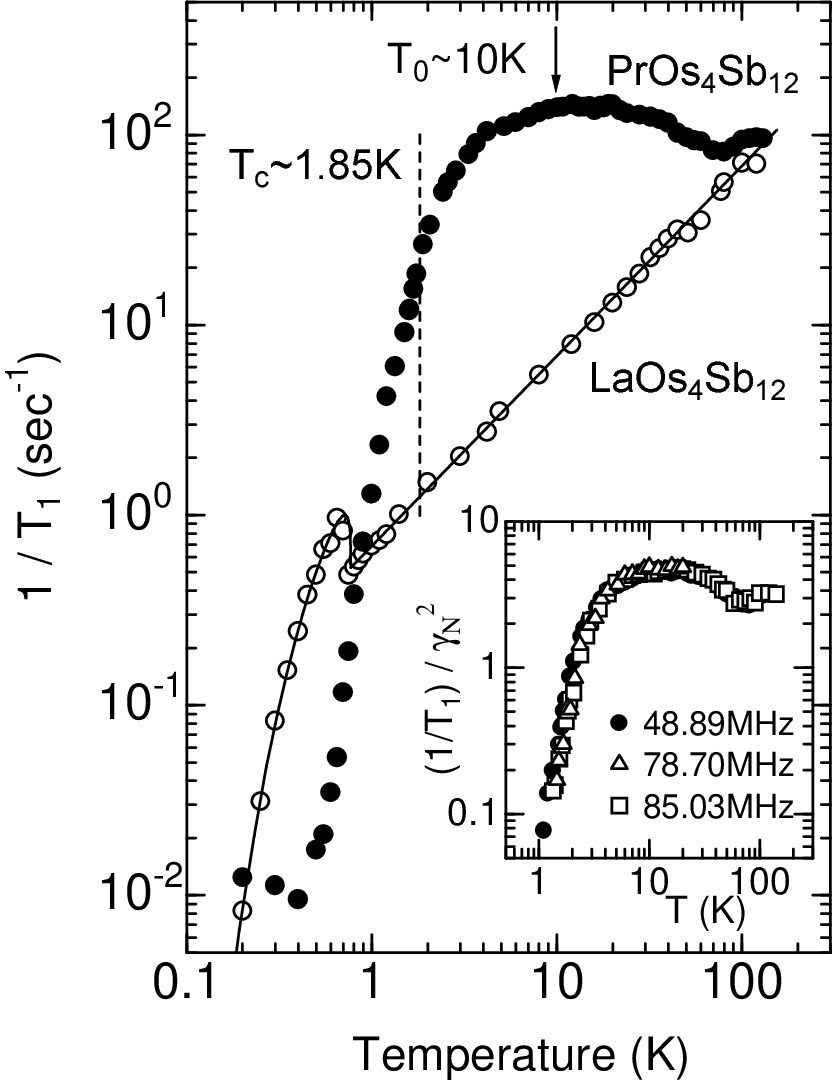}
\caption[]{\protect Temperature dependence of $1/T_1$ at the 2$\nu_Q$ transition of $^{123}$Sb for PrOs$_4$Sb$_{12}$ (closed circle) and LaOs$_4$Sb$_{12}$ (open circle). The inset presents the $T$ dependencies of $1/T_1$'s at the 2$\nu_Q$(48.89 MHz) and 3$\nu_Q$ (78.70 MHz) for $^{123}$Sb and the 2$\nu_Q$ (85.03 MHz) for $^{121}$Sb where the respective data are divided by the nuclear gyromagnetic ratio $^{123}\gamma_N^2$ and $^{121}\gamma_N^2$. All these data are consistent with each other, demonstrating the relaxation process is magnetic in origin.
}
\end{center}
\end{figure}

\clearpage
\begin{figure}[htbp]
\begin{center}
\includegraphics[width=\linewidth]{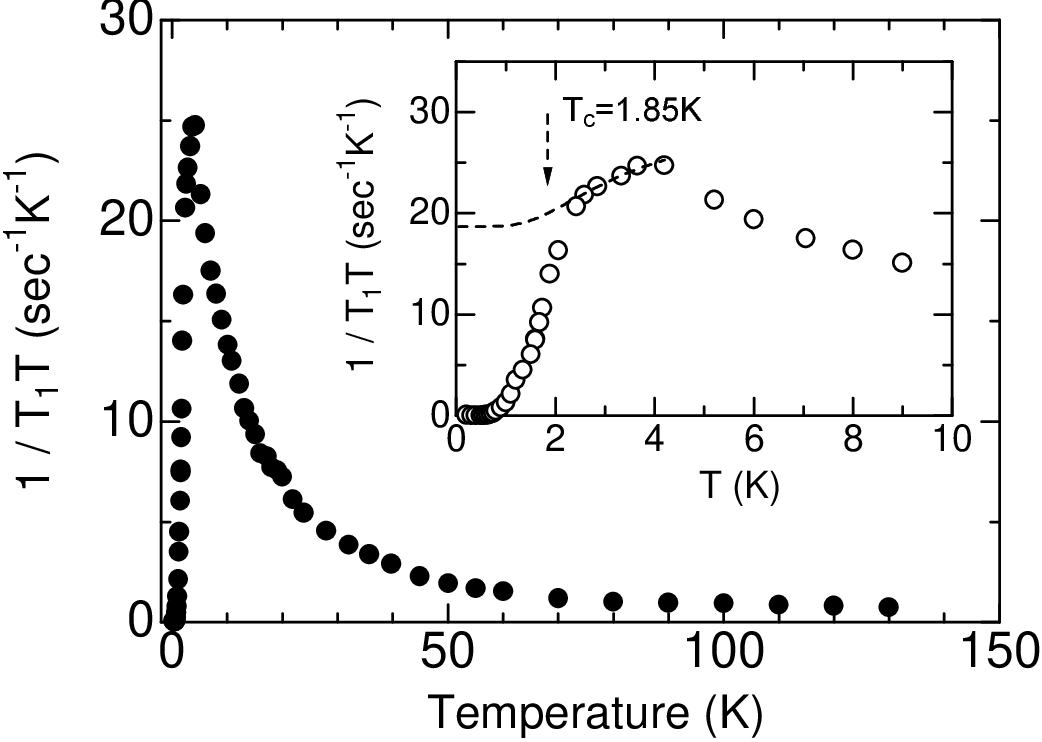}
\caption[]{\protect  $T$ dependence of $1/T_1T$ for PrOs$_4$Sb$_{12}$.  The decrease below $\sim$ 4 K is partially related to the CEF effect. The dotted line in the inset is a calculation based on the model where both the contributions arising from the CEF effect  and the formation of heavy quasipartices are incorporated (see text). The latter is responsible for the onset of the superconductivity.}
\end{center}
\end{figure}

\clearpage
\begin{figure}[htbp]
\begin{center}
\includegraphics[width=\linewidth]{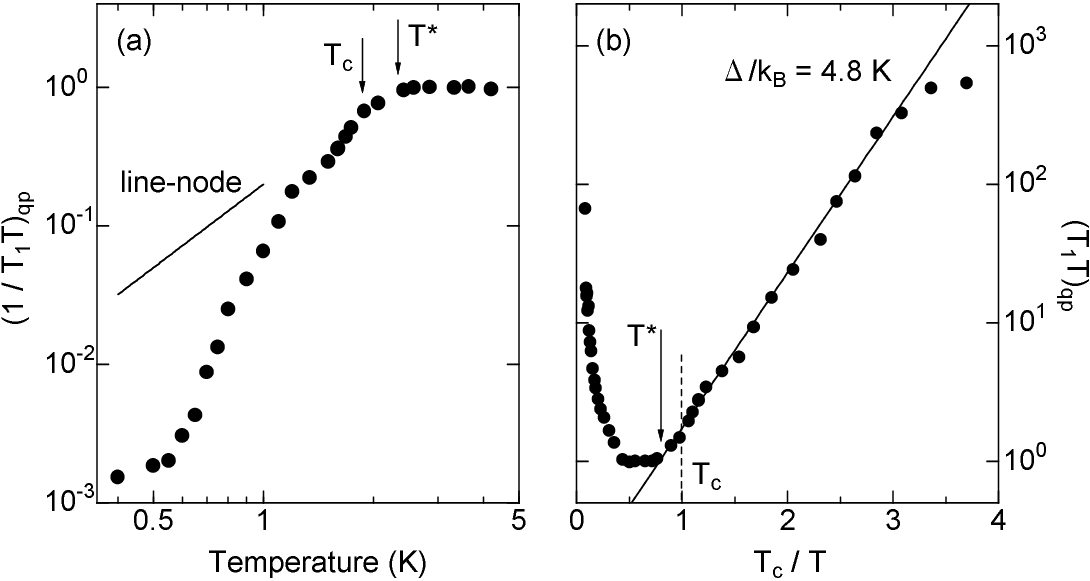}
\caption[]{\protect  $T$ dependence of $(1/T_1T)_{qp}$ corresponding to the heavy quasiparticle contribution. Here the data are normalized by the value of $(1/T_1T)_{qp}$=const. between $T^*=2.3$ K and $T_F\sim$ 4 K. Absence of the coherence peak just below $T_c$=1.85 K and the exponential decrease with $\Delta/k_B=4.8$ K below $T^*\sim$ 2.3 K both highlight that the unconventional strong-coupling superconductivity PrOs$_4$Sb$_{12}$ differs from a conventional s-wave type and any unconventional ones with the line-node gap.
}
\end{center}
\end{figure}

\end{document}